\documentclass[11pt]{article} 
\usepackage{moriond,epsfig,amsmath,amssymb,amsfonts} 

\def\Journal#1#2#3#4{{#1} {\bf #2}, #3 (#4)} 
 
\def\NPB{{\em Nucl. Phys.} B} 
\def\PLB{{\em Phys. Lett.}  B} 
\def\PRL{\em Phys. Rev. Lett.} 
\def\PRD{{\em Phys. Rev.} D} 
\def\ZPC{{\em Z. Phys.} C} 
\def\JHEP{\em JHEP}  
 
\def\mco{\multicolumn}

\def\be{\begin{equation}} 
\def\ee{\end{equation}} 
\def\bea{\begin{eqnarray}} 
\def\eea{\end{eqnarray}} 
 
\begin{document} 

{\begin{flushright}
UG-FT-231/08, CAFPE-101/08
\end{flushright}}
\vspace*{3.75cm} 

\title{ELECTROWEAK CONSTRAINTS ON SEE-SAW MESSENGERS AND THEIR IMPLICATIONS  
FOR LHC} 
 
\author{F. DEL \'AGUILA, J.A. AGUILAR-SAAVEDRA, J. DE BLAS and M. P\'EREZ-VICTORIA} 
 
\address{Departmento de F{\'\i}sica Te\'orica y del Cosmos and CAFPE, \\  
Universidad de Granada, E-18071 Granada, Spain} 

\maketitle\abstracts{ 
We review the present electroweak precision data constraints on the mediators  
of the three types of see-saw mechanisms. Except in the see-saw mechanism of  
type I, with the heavy neutrino singlets being mainly produced through their  
mixing with the Standard Model leptons, LHC will be able to discover or put  
limits on new scalar (see-saw of type II) and lepton (see-saw of type III)  
triplets near the TeV. If discovered, it may be possible in the simplest  
models to measure the light neutrino mass and mixing properties that neutrino  
oscillation experiments are insensitive to.} 
 
\section{Introduction} 
 
\begin{figure}[ht] 
\hspace{2cm}\includegraphics[width=12cm]{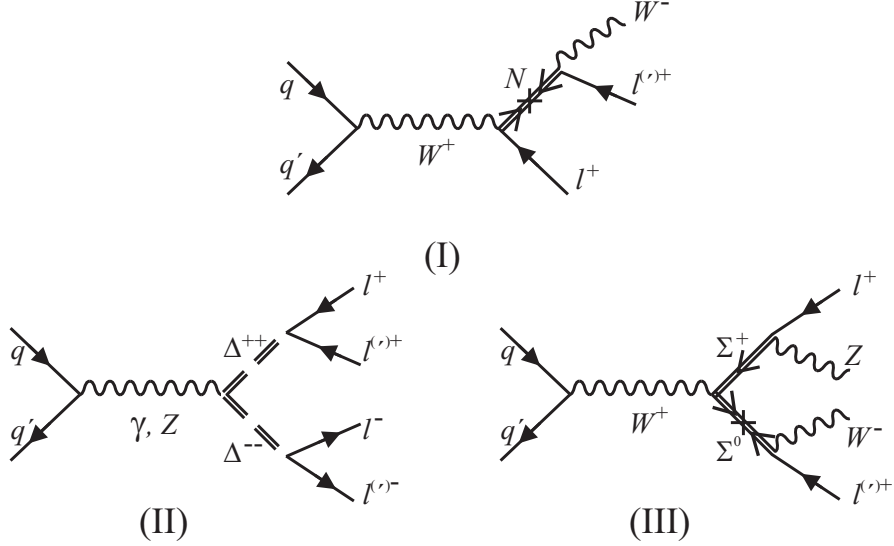} 
\caption{Examples of production diagrams for same-sign dilepton signals,  
$l^+l^{(')+}X$, mediated by the three types of see-saw messengers.  
\label{SSP}}   
\end{figure}

\begin{figure}[ht] 
\hspace{2cm}\includegraphics[width=12cm]{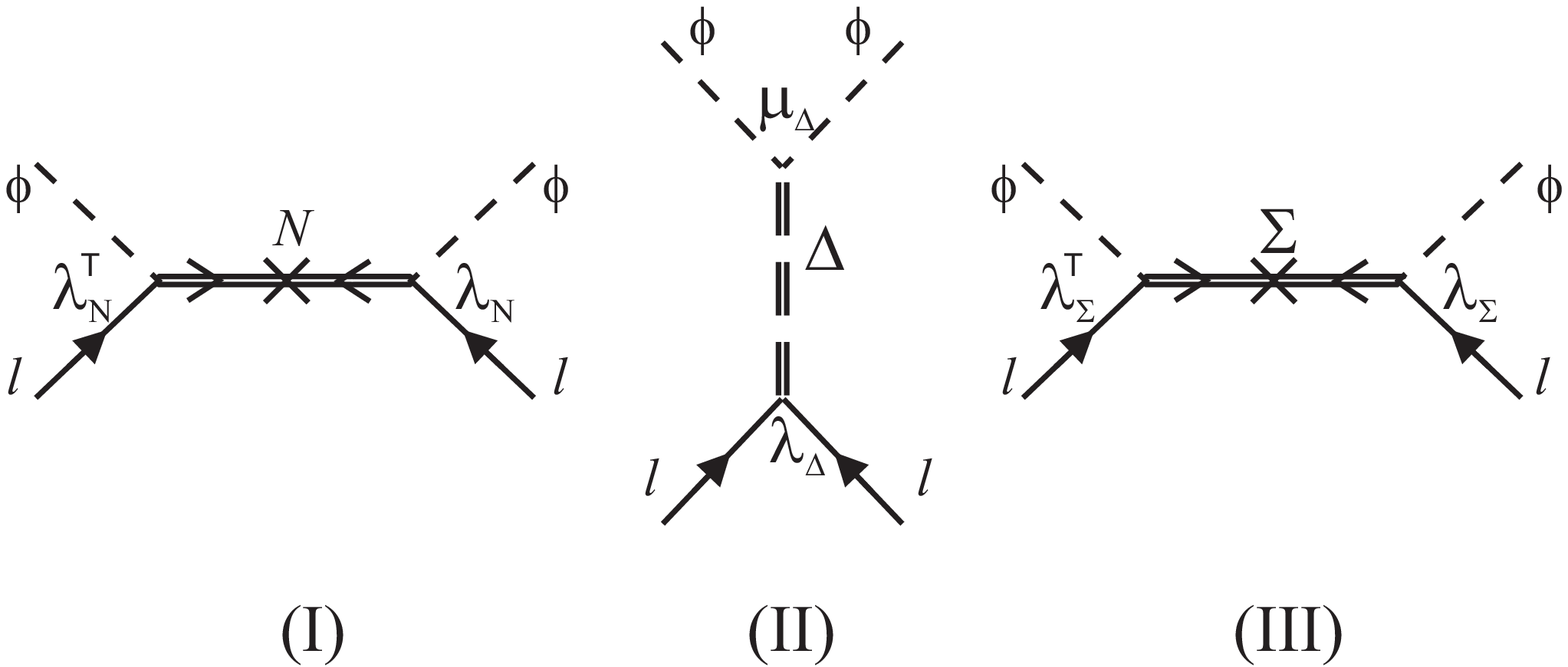} 
\caption{See-saw mechanisms of type I, II and III.~$\lambda_N$, $\lambda_\Delta$  
and $\lambda_\Sigma$ are the Yukawa coupling matrices in the Lagrangian terms  
$-\overline{l_L}\tilde \phi \lambda_N^\dagger N_R$,  
$\overline{\tilde{l}_L} \lambda_\Delta (\vec{\sigma} \cdot \vec{\Delta}) l_L$  
and  
$-\overline{\vec{\Sigma}_R} \lambda_\Sigma (\tilde{\phi}^\dagger \frac{\vec{\sigma}}{2} l_L)$,  
respectively, with $\overline{\tilde {l}_L}=-l_L^TCi\sigma_2$ and $C$  
the spinor charge conjugation matrix. Whereas $\mu_\Delta$ is the coefficient of the  
scalar potential term  
$\tilde \phi^\dagger (\vec{\sigma}\cdot \vec{\Delta})^\dagger \phi$.   
\label{SSM}}   
\end{figure}

As it is well known, the original see-saw mechanism~\cite{See-sawI}, nowadays called of type I, explains the smallness  
of the light neutrino masses $|m_\nu| \sim $ 1 eV invoking a very heavy Majorana  
neutrino $M_N \sim 10^{14}$ GeV:
\begin{equation} 
|m_\nu| \simeq \frac{v^2|\lambda|^2}{M_N} \simeq |V^*|^2 M_N , 
\label{mnu} 
\end{equation} 
where $|\lambda| \sim 1$ is the corresponding Yukawa coupling and $v \simeq 246$ GeV  
the electroweak vacuum expectation value.  
For reviews see \cite{Mohapatra:2005wg,Raidal:2008jk}.  Alternatively, if the heavy scale is at the LHC reach $M_N \sim 1$ TeV, it requires a very small heavy--light mixing angle $|V| \sim 10^{-6}$. 
In its simplest form the model cannot be tested at large colliders, because  
the heavy neutrino $N$ is a Standard Model (SM) singlet and only couples  
to SM gauge bosons through its mixing $V$. Hence it is produced  
through the vertex $-{g}/{\sqrt{2}}\;\overline{\ell}\gamma^\mu V_{\ell N}P_L N W_\mu^-$, 
with $\ell$ a charged lepton, with a cross section
proportional to $|V_{\ell N}|^2$, which is strongly suppressed. See Fig. \ref{SSP}-(I).
%
%
There are two other types of see-saw mechanism giving tree level Majorana masses  
to the light neutrinos $\nu$, as shown in Fig. \ref{SSM}.  
%
%
In all cases the extra particles contribute at low energies to
the dimension 5 lepton number (LN) violating operator~\cite{Weinberg:1979sa}  
\begin{equation} 
({\cal O}_5)_{ij}= \overline{(l_L^i)^c}\tilde \phi ^* \tilde \phi^\dagger l_L^j  
\rightarrow \frac{v^2}{2}\overline{(\nu^i)^c} \nu^j
\hspace{1cm}
( {\rm with}\ l = \left(\begin{array}{c} \nu  \\ \ell \end{array}\right)
\ {\rm and}\ \tilde\phi = i \sigma_2 \phi^*  )\ ,  
\label{Op5}  
\end{equation} 
which gives Majorana masses to light neutrinos after spontaneous symmetry breaking.
The see-saw of type II ~\cite{See-sawII} in Fig. \ref{SSM} is mediated by an 
$SU(2)_L$ scalar triplet $\Delta$ of hypercharge $Y = 1$, 
implying three new complex scalars of charges  
$Q = T_3 + Y$: $\Delta^{++}, \Delta^+, \Delta^0$.  
The see-saw of type III ~\cite{See-sawIII} exchanges an $SU(2)_L$ 
fermion triplet $\Sigma$ of hypercharge $Y = 0$, assumed to be Majorana and 
containing charged leptons $\Sigma^{\pm}$ and a Majorana neutrino $\Sigma^0$.  
The main difference for LHC detection is that the see-saw messengers for  
these last two mechanisms can be produced by unsuppressed processes of electroweak size (Fig. \ref{SSP}). Their decay, even if suppressed by small couplings,  
can take place within the detector due to the large mass of the  
new particle. 
All three types of see-saw messengers produce LN conserving as well as LN violating 
signals, but the former have much larger backgrounds. 
On the other hand, same-sign dilepton signals, $l^\pm l^{(')\pm} X$, do not have to be 
necessarily LN violating. 
Thus, in the example in Fig. \ref{SSP}--(II), the decay coupling 
$\lambda_\Delta$ needs not be very small because it is only one of the factors  
entering in the LN violating expression for $\nu$ masses (see Table \ref{tab:coe}).  
\begin{table}[thb] 
\caption{Coefficients of the operators up to dimension 6  
arising from the integration of the heavy  
fields involved in each see-saw model. The parameters $\lambda_3$ and $\lambda_5$  
are the coefficients of the scalar potential terms  
$-(\phi^\dagger \phi)(\vec{\Delta}^\dagger \vec{\Delta})$  
and  
$-(\vec{\Delta}^\dagger T_i \vec{\Delta})(\phi^\dagger \sigma_i \phi)$,  
respectively, 
and $(\lambda_e)_{jj}$ the diagonalised SM charged-lepton Yukawa couplings. 
The remaining parameters are defined in the caption of Fig. \ref{SSM}. 
\vspace{0.4cm} 
\label{tab:coe}} 
\begin{center} 
\begin{tabular}{|c|c|c|c|} 
\hline 
Coefficient & Type I & Type II & Type III \\ 
\hline 
$ $&$ $&$ $&$ $\\[-.4cm] 
$\alpha_4$ & $-$ &$2\frac{\left|\mu_\Delta\right|^2}{M_\Delta^2} $ & $-$ \\[.2cm] 
\hline 
$ $&$ $&$ $&$ $\\[-.33cm] 
$\frac{(\alpha_5)_{ij}}{\Lambda}$ &$\frac 12 \frac{(\lambda^T_{N})_{ia} 
(\lambda_{N})_{aj}}{M_{N a}}$ &$-2\frac{\mu_\Delta(\lambda_{\Delta})_{ij}}
{M_{\Delta}^2}$&$\frac 18 \frac{(\lambda^T_{\Sigma})_{ia} 
(\lambda_{\Sigma})_{aj}}{M_{\Sigma a}}$ \\[.25cm] 
\hline 
$ $&$ $&$ $&$ $\\[-.35cm] 
$\frac{(\alpha_{\phi l}^{(1)})_{ij}}{\Lambda^2}$ &$\frac 14 \frac{ 
(\lambda^\dagger_{N})_{ia}(\lambda_{N})_{aj}}{M_{N a}^2}$  
& $-$ &$\frac {3}{16} \frac{(\lambda^\dagger_{\Sigma})_{ia} 
(\lambda_{\Sigma})_{aj}}{M_{\Sigma a}^2}$ \\[.2cm] 
$\frac{(\alpha_{\phi l}^{(3)})_{ij}}{\Lambda^2}$ &$-\frac{ 
(\alpha_{\phi l}^{(1)})_{ij}}{\Lambda^2}$ & $-$ &$\frac 13\frac{ 
(\alpha_{\phi l}^{(1)})_{ij}}{\Lambda^2}$ \\[.2cm] 
$\frac{(\alpha_{ll}^{(1)})_{ijkl}}{\Lambda^2}$ & $-$ &$2\frac{ 
(\lambda_\Delta)_{jl}(\lambda^\dagger_\Delta)_{ki}}
{M_\Delta^2}$ & $-$ \\[.2cm] 
$\frac{\alpha_\phi}{\Lambda^2}$ & $-$ &$-6(\lambda_3+\lambda_5) 
\frac{\left|\mu_\Delta\right|^2}{M_\Delta^4}$ & $-$ \\[.2cm] 
$\frac{\alpha_\phi^{(1)}}{\Lambda^2}$ & $-$ &$4\frac{\left|\mu_\Delta 
\right|^2}{M_\Delta^4}$ & $-$ \\[.2cm] 
$\frac{\alpha_\phi^{(3)}}{\Lambda^2}$ & $-$ &$4\frac{\left|\mu_\Delta 
\right|^2}{M_\Delta^4}$ & $-$ \\[.2cm] 
$\frac{(\alpha_{e \phi})_{ij}}{\Lambda^2}$ & & $-$ &$\frac 4 3\frac{ 
(\alpha_{\phi l}^{(1)})_{ij}}{\Lambda^2}(\lambda_e)_{jj}$ \\[.2cm] 
\hline 
\end{tabular} 
\end{center} 
\end{table} 
In fact, this process is LN conserving as we can conventionally assign LN 
equal to 2 to $\Delta^{--}$. 
There are other processes that do violate LN, e.g. when one of the 
doubly-charged $\Delta$ in Fig. \ref{SSM}--(II) decays into $WW$. 
Then, what does violate LN is the corresponding $\Delta WW$ vertex, which is 
proportional to the coupling of the only LN violating term 
in the fundamental Lagrangian 
$\tilde \phi^\dagger(\vec{\sigma}\cdot \vec{\Delta})^\dagger \phi$, 
with total LN equal to 2. 
In the examples in Fig. \ref{SSP}--(I, III) LN is violated in the decay (mass) 
of the heavy neutral fermion.  
 
In conclusion, all the three mechanisms produce same-sign dilepton  
signals, but only the last two are observable at LHC  
\cite{Han:2006ip,del Aguila:2007em,Gunion:1989ci,Akeroyd:2005gt,Hektor:2007uu,Perez:2008zc,Franceschini:2008pz} 
in minimal models. Heavy neutrino singlets in particular non-minimal scenarios could 
also be observed, as described in Section~\ref{Signals}.
 
In the following we first review the experimental constraints on the  
parameters entering the three see-saw mechanisms, and then the LHC reach for the corresponding see-saw messengers.
Complementary reviews on this subject have been presented by other speakers  
at this Conference (see F. Bonnet, T. Hambye and J. Kersten in these Proceedings). 
 
\section{Electroweak precision data limits on see-saw messengers} 
 
The low energy effects of the see-saw messengers can be described by the  
effective Lagrangian 
\begin{equation} 
{\cal L}_\mathrm{eff}={\cal L}_4+\frac{1}{\Lambda}{\cal 
  L}_5+\frac{1}{\Lambda^2}{\cal L}_6+\dots  , 
\label{Leff} 
\end{equation} 
where $\Lambda$ is the cut-off scale, in our case of  
the order of the see-saw messenger masses $M$, and  
the different terms contain gauge-invariant operators of the corresponding dimension.  
The non-zero terms up to dimension 6 are  
\cite{delAguila:2007ap,Abada:2007ux} 
\begin{equation} 
{\cal L}_4={\cal L}_{SM}+\alpha_4\left(\phi^\dagger \phi\right)^2 ,  
\label{L4}  
\end{equation} 
\begin{equation} 
{\cal L}_5=(\alpha_5)_{ij} \overline{(l_L^i)^c}\tilde \phi ^* \tilde 
\phi^\dagger l_L^j+\mbox{h.c.} , 
\label{L5}  
\end{equation} 
\begin{equation} 
\begin{split} 
{\cal L}_6=&\left[(\alpha_{\phi l}^{(1)})_{ij}  
\left(\phi^\dagger iD_\mu \phi\right)\left(\overline{l_L^i}\gamma^\mu l_L^j\right)  
+ (\alpha_{\phi l}^{(3)})_{ij} \left(\phi^\dagger 
i\sigma_aD_\mu \phi\right)\left(\overline{l_L^i}\sigma_a\gamma^\mu 
l_L^j\right)\right.\\  
&\left.+\left(\alpha_{e\phi}^{ }\right)_{ij} 
\left(\phi^\dagger\phi\right)\left(\overline{l_L^i}\phi e_R^j\right)+ 
(\alpha_{ll}^{(1)})_{ijkl}\frac 1 2 \left( 
\overline{l_L^i}\gamma^\mu l_L^j\right)\left(\overline{l_L^k} 
\gamma_\mu l_L^l\right)+\mbox{h.c.}\right]\\ 
&+\alpha_{\phi}^{(1)} \left(\phi^{\dagger}\phi\right) 
\left(\left(D_\mu \phi\right)^\dagger D^\mu \phi\right)+\alpha_{\phi}^{(3)} 
\left(\phi^{\dagger}D_\mu \phi\right)\left( (D^\mu \phi)^\dagger  
\phi\right)+\alpha_{\phi} \frac 1 3 \left(\phi^\dagger \phi\right)^3 ,  
\label{L6}  
\end{split}  
\end{equation} 
where we choose the basis of B\"uchmuller and Wyler to express the  
result \cite{Buchmuller:1985jz}. $l_L$ stands for any  
lepton doublet, $e_R$ for any lepton singlet, and $\phi$ is the  
SM Higgs doublet.  
In Table \ref{tab:coe} we collect the explicit expressions of the  
coefficients in terms of the original parameters for each type of see-saw 
(see Fig. \ref{SSM} and the table caption for definitions).  

Only the dimension 6 operators can give deviations from the SM predictions  
for the electroweak precision data (EWPD). 
The operators of dimension 4 only redefine SM parameters. 
The one of dimension 5 gives tiny masses to the light neutrinos, 
and contributes to neutrinoless double $\beta$ decay.  
An important difference is that the coefficient $\alpha_5$ involves LN-violating 
products of two $\lambda$'s or of $\mu$ and $\lambda$, while the other coefficients 
depend on $\lambda^* \lambda$ or $|\mu|^2$. Therefore, it is possible to have large 
cancellations in $\alpha_5$ together with sizeable coefficients of 
dimension six~\cite{delAguila:2007ap,Abada:2007ux}. 
Type I and III fermions generate
the operators ${\cal O}^{(1,3)}_{\phi l}$, which correct the gauge fermion  
couplings. Type II scalars, on the other hand, generate 4-lepton operators and the 
operator ${\cal O}^{(3)}_{\phi}$, which breaks custodial symmetry and modifies the 
SM relation between the gauge boson masses. EWPD are sensitive to all these effects 
and put limits on the see-saw parameters. 

There are two classes of processes, depending on  
whether they involve neutral currents violating lepton flavour (LF) or not.  
The first class puts more stringent limits
\cite{Antusch:2006vwa,Abada:2008ea}, but only on the combinations 
of coefficients entering off-diagonal elements. The second class is measured 
mainly at LEP~\cite{Yao:2006px} and constrains the combinations in the diagonal 
entries~\cite{delAguila:2008pw}. The LF violating limits are satisfied in types 
I and III if $N$ and $\Sigma$ mainly mix with only one charged lepton family.  
In Table \ref{LeptLimits} we collect the bounds from EWPD on the  
$N$ and $\Sigma$ mixings with the SM leptons $V_{\ell N,\ell \Sigma}$ 
\cite{delAguila:2008pw}, and in  
Table \ref{LFLimits} their product including the LF violating  
bounds \cite{Antusch:2006vwa,Abada:2008ea}.  
\begin{table}[ht] 
\caption{Upper limit at 90 $\%$ confidence level (CL) on the absolute value of the  
mixings. The first 
three columns are obtained by coupling each new lepton with only one SM 
family. The last one corresponds to the case of lepton 
universality: three new lepton multiplets mixing with only one  
charged-lepton family each, all of them with the same mixing angle.  
All numbers are computed assuming $M_H \geq 114.4~\mbox{GeV}$. 
\label{LeptLimits}
\vspace{0.4cm}}   
\begin{center} 
\begin{tabular}{| c | c c c c|}\hline 
$ $&$ $&$ $&$ $&$ $\\[-.35cm] 
$\mbox{Coupling} $&$\mbox{Only with }e$&$\mbox{Only with }\mu $ 
&$\mbox{Only with }\tau $&$\mbox{Universal} $\\[.1cm] 
\hline 
$ $&$ $&$ $&$ $&$ $\\[-.35cm] 
$\left|V_{\ell N}=\frac{v(\lambda^\dagger_N)_{l N}}
{\sqrt{2}M_N}\right|<$&$0.055 $ 
&$0.057 $&$0.079 $&$0.038$\\[.1cm] 
$ $&$ $&$ $&$ $&$ $\\[-.25cm] 
$\left|V_{\ell \Sigma}=-\frac{v(\lambda^\dagger_\Sigma)_{l \Sigma}}
{2\sqrt{2}M_\Sigma}\right|<$ 
&$0.019$&$0.017 $&$0.027 $&$0.016 $\\[.25cm]  
\hline  
\end{tabular} 
\end{center} 
\end{table} 
\begin{table}[ht] 
\caption{Upper limit at 90 $\%$ CL on the absolute value of the  
products of the mixings between heavy singlets $N$ and triplets 
$\Sigma$ with the SM leptons, $VV^*$, entering in low energy processes.  Row and column ordering corresponds to $e, \mu , \tau$.
\label{LFLimits} 
\vspace{0.4cm}} 
\begin{center} 
\begin{tabular}{|ccc|ccc|} 
\hline 
$ $&$ $&$ $&$ $&$ $&$ $\\[-.35cm] 
\mco{3}{|c|}{$|V_{\ell N}V^*_{\ell 'N}| <$}  
& \mco{3}{|c|}{$|V_{\ell \Sigma}V^*_{\ell '\Sigma}| <$} \\[.1cm] 
\hline 
$ $&$ $&$ $&$ $&$ $&$ $\\[-.3cm] 
0.0030 & 0.0001 & 0.01 & 0.0004 &  
$1.1 \times 10^{-6}$ & 0.0005 \\ 
0.0001 & 0.0032 & 0.01 &  
$1.1 \times 10^{-6}$ & 0.0003 & 
0.0005 \\ 
0.01 & 0.01 & 0.0062 & 0.0005 &  
0.0005 & 0.0007 \\ 
\hline 
\end{tabular} 
\end{center} 
\end{table} 
These values update and extend previous bounds on diagonal entries for $N$ 
\cite{Bergmann:1998rg,Tommasini:1995ii} 
(see also \cite{Langacker:1988ur}.) 
Their dependence on the model parameters entering in the operator  
coefficients in Table \ref{tab:coe} is explicit in the first column  
of Table \ref{LeptLimits}.   
All low energy effects are proportional to this mixing, and the same holds for   
the gauge and Higgs couplings between the new and the SM leptons,
responsible of the heavy lepton decay (and $N$ production if there  
is no extra NP).   
An interesting by-product of a non-negligible mixing of the electron or  
muon with a heavy $N$ is that the fit to EWPD prefers a Higgs mass $M_H$ higher than in the SM,  
in better agreement with the present direct limit. This is so because  
their contributions to the most significative observables partially cancel~\cite{Loinaz:2002ep}, so that both the mixing and $M_H$ can be relatively large without spoiling the agreement with EWPD.  
The new 90 \% CL on $M_H$ increases in this case \cite{delAguila:2008pw} 
up to $\sim$ 260 GeV  
(see also \cite{Cynolter:2008ea,Gogoladze:2008ak}).  
In all other cases the limit stays at $\sim 165$ GeV. 
 
In type II see-saw a crucial phenomenological issue is the relative size  
of $(\lambda _\Delta)_{ij}$ and $\mu _\Delta$ for $M_\Delta \sim 1$ TeV.  
The $\nu$ masses are proportional to their product, 
$(m_\nu)_{ij} = 2v^2\frac{\mu_{\Delta}(\lambda_\Delta)_ {ij}}{M_\Delta^2}$, 
which gives the strength of the LN violation.  
If $\mu _\Delta$ is small enough, $(\lambda _\Delta)_{ij}$ can be  
relatively large and saturate present limits on LF violating processes, 
eventually showing at the next generation of experiments.  
If instead $(\lambda _\Delta)_{ij}$ are very small, the flavour structure appears only in the $\nu$ mass matrix.  
The present limits are reviewed in \cite{Abada:2007ux}. Neglecting LF  
violating bounds (i.e., assuming that $(\lambda _\Delta)_{ee}$ is small enough  
not to give a too large $\mu \rightarrow e \bar e e$ decay rate), $\mu_\Delta$ and $\lambda_\Delta$ are constrained by the $T$ oblique parameter and four-fermion processes, respectively. From a global fit to EWPD (see \cite{delAguila:2008pw} for details on the  
data set used) we obtain the following limits at 90 $\%$ CL:
\begin{eqnarray} 
\frac{\left|\mu_\Delta\right|}{M_\Delta^2}<0.048\mbox{TeV}^{-1}, 
&\frac{\left|(\lambda_\Delta)_{e\mu}\right|}{M_\Delta}<0.100\mbox{TeV}^{-1} .  
\label{limit}
\end{eqnarray} 

\section{Dilepton signals of see-saw messengers 
\label{Signals}}

The previous limits apply to any particle transforming as the corresponding  
see-saw messenger, independently of whether it contributes or not to light  
neutrino masses. As indicated above, in minimal models the tight restriction 
imposed by $\nu$ masses (Eq. \ref{mnu}) gives much more stringent limits for the 
mixings of TeV-scale see-saw messengers. However, these limits can be avoided if 
additional particles give additional contributions to neutrino masses that 
cancel the previous ones, for instance if the fermionic messengers are 
quasi-Dirac, i.e. a nearly degenerate Majorana pair with appropriate couplings 
\cite{Wyler:1982dd}. 
The EWPD limits are in this case relevant for production and detection of type I 
messengers $N$, but the signals are different because they conserve LN to a 
very large extent ~\cite{delAguila:2007ap,Kersten:2007vk}. 
On the other hand, type II and III messengers with  
masses near the TeV can be produced and detected at LHC even in minimal models. 
Let us discuss the three types of see-saw mechanism in turn. 

\subsection{Type I: Fermion singlets $N$} 

As already explained, a type I heavy  
neutrino $N$ with a mixing saturating the EWPD limit cannot be Majorana,  
unless extra fields with a very precise fine tuning keep the $\nu$   
masses small enough~\cite{Ingelman:1993ve}. 
Unnatural cancellations allowing for LN-violating signals are also possible 
in principle. In this case a
fast simulation shows that LHC can discover a Majorana neutrino singlet  
with $M_N \simeq 150$ GeV for $|V_{\mu N}| \geq 0.054$ (near the EWPD limit) 
\cite{del Aguila:2007em}, 
assuming an integrated luminosity ${\sf L} = 30\ fb^{-1}$.

Such a signal can be also observed for much smaller mixings and  
larger masses if there is some extra NP \cite{delAguila:2008}, 
especially if the extra particles can be copiously produced at LHC 
\cite{delAguila:2008iz}.  
This is the case, for instance, if the gauge group is left-right symmetric  
and the new $W'_R$ has a few TeV mass.  
Then $pp \rightarrow W'_R \rightarrow \ell N \rightarrow \ell \ell'W$ is observable, 
even with negligible mixing $V_{\ell N}$, for  
$M_N $ and $M_{W'_R}$ up to 2.3 TeV and 3.5 TeV, respectively,  
\cite{Gninenko:2006br} for an integrated luminosity ${\sf L} = 30\ fb^{-1}$. 
Similarly, if the SM is extended with a leptophobic $Z'$, the process  
$pp \rightarrow Z' \rightarrow NN \rightarrow \ell \ell'WW$   
can probe $Z'$ masses~\cite{delAguila:2007ua} up to 2.5 TeV, and $M_N$ 
up to 800 GeV. 

\subsection{Type II: Scalar triplets $\Delta$} 

\begin{figure}[ht] 
\hspace{3cm}\includegraphics[width=9.5cm]{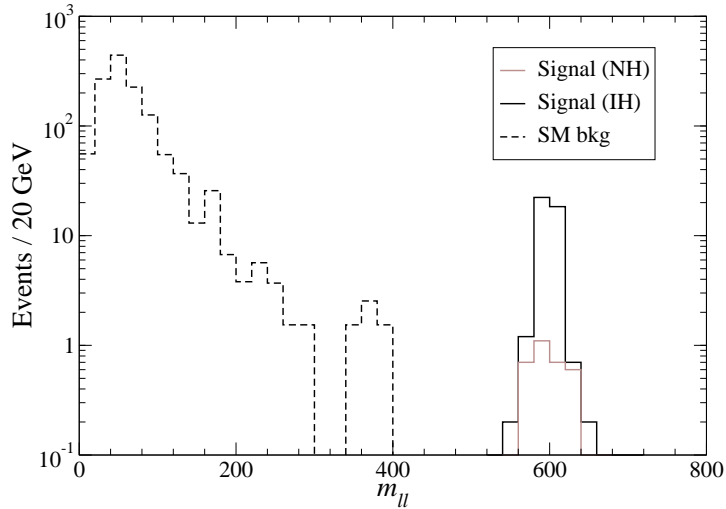} 
\caption{Same-sign dilepton invariant mass distributions for 
$M_\Delta = 600$ GeV and normal (NH) and inverted (IH) $\nu$ 
mass hierarchies, assuming an integrated luminosity ${\sf L} = 300\ fb^{-1}$. 
\label{Mll}}   
\end{figure}

$SU(2)_L$ scalar triplets can be produced through the exchange of electroweak  
gauge bosons with SM couplings, and then they may be observable for masses near  
the TeV scale (see for reviews \cite{Raidal:2008jk,delAguila:2008iz}).   
Although suppressed, their decays can occur within the detector  
for these large masses. In Fig. \ref{SSP}-(II) we display one of the possible processes.  
The search strategy and LHC potential depend on the dominant decay 
modes. These are proportional to the $\Delta$ vacuum expectation value 
$|< \Delta^0 >| \equiv v_\Delta$, as for example \cite{Gunion:1989ci}
$\Delta ^{\pm\pm} \rightarrow W^{\pm}W^{\pm}$, 
or to $(\lambda_\Delta)_{ij}$, as \cite{Hektor:2007uu}
$\Delta ^{\pm\pm} \rightarrow l^{\pm}l^{(')\pm}$.
$\Delta ^{\pm\pm}$ can also decay into 
$\Delta^{\pm}W^{\pm *}$ if kinematically allowed 
(see \cite{Akeroyd:2005gt}). 
All these different decay channels make the phenomenological 
analysis of single and pair $\Delta^{\pm\pm}$ production quite rich 
\cite{Perez:2008zc}.
The EWPD limit in Eq. \ref{limit} translates into the bound
$v_\Delta = \frac{v^2|\mu_{\Delta}|}{\sqrt 2M_\Delta^2} < 2$ GeV. 
This is to be compared with  
$|m_\nu| = 2\sqrt 2 v_\Delta |\lambda_\Delta| \sim 10^{-9}$ GeV, which gives a much 
more stringent constraint for  non-negligible $\lambda_\Delta$. 
Dilepton (diboson) decays are dominant for  
$v_\Delta < (>)\ v^c_\Delta \sim 10^{-4}$ GeV. If for instance  
$\lambda_\Delta$ is of the same size as the charged lepton Yukawa 
couplings $\sim 10^{-2} - 5 \times 10^{-6}$, $v_\Delta$ varies from $5\times 10^{-8}$ 
to $10^{-4}$ GeV, below the critical value $v_\Delta^c$, and $\Delta$ 
decays mainly into leptons. In this case the LHC reach for 
$M_{\Delta^{\pm\pm}}$ has been estimated, based on statistics, to be $\sim$ 1 TeV 
for an integrated luminosity ${\sf L} = 300\ fb^{-1}$.  
In Fig. \ref{Mll} we plot 
the invariant mass distribution $m_{\ell \ell}$ of same-sign 
dilepton pairs containing the lepton of largest transverse momentum 
for $M_\Delta = 600$ GeV. 
As this fast simulation analysis shows, the 
SM background is well separated from the signal, 
and the LHC discovey potential
strongly depends on the light neutrino mass hierarchy.
For the simulated sample we find 4 (44) signal events for 
the normal $\nu$ mass hierarchy NH (inverted IH), well 
separated from the main backgrounds:  
$t\bar t nj$ (1007 events), $Zb\bar b nj$ (91 events), 
$tW$ (68 events), and $Zt\bar t nj$ (51 events). 
We get rid of other possible backgrounds like $ZZ nj$ 
requiring no opposite-sign dilepton pairs with an 
invariant mass in the range $M_Z \pm 5$ GeV.
For larger $v_\Delta$ values, with dominant non-leptonic 
decays, the corresponding reach estimate based on statistics 
is $\sim$ 600 GeV. 
Note that only in the leptonic case LHC is sensitive to the see-saw 
flavour structure. Near the critical value, one could in principle 
extract information on the structure and on the global scale of the see-saw. 

Tevatron Collaborations have already established limits on 
the scalar triplet mass assuming that $\Delta^{\pm\pm} \rightarrow l^{\pm}l^{\pm}$ 
100 $\%$ of the time:   
At the 95 $\%$ CL $M_{\Delta^{\pm\pm}} > 150$ GeV for 
$\Delta^{\pm\pm}$ only decaying to muons \cite{:2008iy}, 
and an integrated luminosity ${\sf L} = 1.1\ fb^{-1}$.  
%
%

\subsection{Type III: Fermion triplets $\Sigma$} 

Not so much attention has been payed to the study of the LHC reach 
for $SU(2)_L$ fermion triplets $\Sigma$. Up to very recently a similar electroweak 
process, the production of a heavy 
vector-like lepton doublet \cite{delAguila:1989rq}, 
had to be used to guess that LHC could be sensitive 
to $M_\Sigma \sim 500$ GeV. A dedicated study 
\cite{Franceschini:2008pz} estimates that an integrated 
luminosity ${\sf L} = 10\ fb^{-1}$ should allow to observe LN violating 
signals (see Fig. \ref{SSP}-(III) for a relevant process) for 
$M_\Sigma < 800$ GeV.  
Vector-like fermion triplets couple to SM leptons proportionally to 
its mixing $V_{l\Sigma}$, which is $\leq 10^{-6}$ according to Eq. \ref{mnu} 
if $\Sigma$ is at the LHC reach $\sim 1$ TeV. 
So, one can eventually improve the analysis using the displaced vertex 
signatures of their decays. 
 
\section{Conclusions} 
 
Same-sign dilepton signals $l^\pm l(')^\pm X$ will allow to set significative 
limits on see-saw messengers at LHC, 
as illustrated in Table \ref{SeesawLimits}.
\begin{table}[ht] 
\caption{LHC discovery limit estimates for see-saw messengers, assuming  
an integrated luminosity ${\sf L} = 30, 300\ {\rm and}\ 10\ fb^{-1}$ for 
$N,~\Delta~{\rm and}\ \Sigma$, 
respectively. See Section \ref{Signals} for a detailed explanation.
\label{SeesawLimits}} 
\vspace{0.4cm} 
\begin{center} 
\begin{tabular}{|c|c|c|c|} 
\hline 
$ $&$ $&$ $&$ $\\[-.35cm] 
 & $M_N$ & $M_\Delta$ & $M_\Sigma$ \\[.1cm] 
\hline 
$ $&$ $&$ $&$ $\\[-.3cm] 
LHC reach (in GeV) & 150 & $600 - 1000$ & 800 \\[.1cm] 
\hline 
\end{tabular} 
\end{center} 
\end{table} 
The estimates for $M_\Delta$ and $M_\Sigma$ are mainly based on statistics, 
and a more detailed analysis is needed to confirm them. 

\section*{Acknowledgments} 
 
We thank Ll. Ametller, 
S. Bar-Shalom, C. Biggio, T. Hambye, A. Soni and J. Wudka for  
discussions. 
F.A. thanks the organizers of the Rencontres de Moriond EW 2008 meeting  
for the excellent organization and the warm hospitality.  
This work has been supported by MEC project FPA2006-05294 and Junta de  
Andaluc{\'\i}a projects FQM 101, FQM 437 and FQM03048.  
J.A.A.S. and J.B. also thank MEC for a Ram\'on y Cajal and an FPU grant, 
respectively.

\end{document}